\documentclass[lettersize,journal]{IEEEtran}
\usepackage{amsmath,amsfonts}
\usepackage{algorithmic}
\usepackage{algorithm}
\usepackage{array}
\usepackage[caption=false,font=normalsize,labelfont=sf,textfont=sf]{subfig}
\usepackage{textcomp}
\usepackage{stfloats}
\usepackage{url}
\usepackage{verbatim}
\usepackage{graphicx}
\usepackage{cite}
\usepackage[normalem]{ulem}
\usepackage[dvipsnames]{xcolor}

\usepackage{booktabs}
\usepackage{enumitem}
\usepackage{multirow}
\usepackage{xcolor}
\DeclareMathAlphabet{\pazocal}{OMS}{zplm}{m}{n}
\usepackage{colortbl}
\definecolor{Scolor}{HTML}{B04C7A}
\definecolor{Mcolor}{HTML}{E06F85}
\definecolor{Wcolor}{HTML}{E79C82}

\begin{document}

\title{Tricolore: Multi-Behavior User Profiling for Enhanced Candidate Generation in Recommender Systems}

\author{Xiao Zhou, Zhongxiang Zhao, and Hanze Guo
\thanks{\it{Xiao Zhou and Hanze Guo are with Gaoling School of Artificial Intelligence, Renmin University of China, Beijing 100872, China. Email:xiaozhou@ruc.edu.cn,ghz@ruc.edu.cn.}}
\thanks{\it{Zhongxiang Zhao is with WeChat Business Group, Tencent Inc. Beijing, China. Email:abnerzxzhao@tencent.com.}}
\thanks{\it{Corresponding author: Xiao Zhou.}}}

\markboth{Journal of IEEE Transactions on Knowledge and Data Engineering}%
{Shell \MakeLowercase{\textit{et al.}}: A Sample Article Using IEEEtran.cls for IEEE Journals}


\maketitle

\begin{abstract}

Online platforms aggregate extensive user feedback across diverse behaviors, providing a rich source for enhancing user engagement. Traditional recommender systems, however, typically optimize for a single target behavior and represent user preferences with a single vector, limiting their ability to handle multiple important behaviors or optimization objectives. This conventional approach also struggles to capture the full spectrum of user interests, resulting in a narrow item pool during candidate generation. To address these limitations, we present \textit{Tricolore}, a versatile multi-vector learning framework that uncovers connections between different behavior types for more robust candidate generation. \textit{Tricolore}'s adaptive multi-task structure is also customizable to specific platform needs. To manage the variability in sparsity across behavior types, we incorporate a behavior-wise multi-view fusion module that dynamically enhances learning. Moreover, a popularity-balanced strategy ensures the recommendation list balances accuracy with item popularity, fostering diversity and improving overall performance. Extensive experiments on public datasets demonstrate \textit{Tricolore}'s effectiveness across various recommendation scenarios, from short video platforms to e-commerce. By leveraging a shared base embedding strategy, \textit{Tricolore} also significantly improves the performance for cold-start users. The source code is publicly available at: \url{https://github.com/abnering/Tricolore}.

\end{abstract}


\section{Introduction}

\IEEEPARstart{W}{ith} the mission of delivering personalized recommendations to a massive user base and addressing information overload in the digital era, recommender systems have become integral to various platforms~\cite{tang2016empirical,xia2021graph,huang2019online,shi2018heterogeneous,ricci2011introduction,li2024ada,ma2024tail}. Despite notable progress, existing recommendation techniques predominantly focus on optimizing a single type of interaction, often tied directly to platform profitability, such as \textit{purchase} in e-commerce, \textit{rating} in movie recommendations, and \textit{click} feedback for online news services~\cite{tang2016empirical,jin2020multi}. Unfortunately, these target behaviors, being high-cost and low-frequency digital traces, lead to cold-start and data sparsity issues, inaccurate user preference representations, and substantial performance degradation~\cite{jin2020multi,wu2022multi,wu2023m2eu}. To address this, some studies~\cite{zhou2019collaborative,gao2019neural,jin2020multi} have explored diverse user feedback types, giving rise to the emerging field of multi-behavior recommender systems (MBRS). These supplementary user behaviors, such as \textit{browse, click, share} for news sites, \textit{listen, favorite, add to playlist} on music platforms, and \textit{view, click, add to cart} for online marketplaces, are often relatively richer and more readily available across domains. The core concept of MBRS is leveraging these multiple feedback types to learn more accurate user preferences for high-quality recommendations~\cite{wu2022multi}. As a nascent field within recommender systems, there is ample room for enhancement.

\begin{figure}[t]
    \centering
    \includegraphics[width=0.4\textwidth]{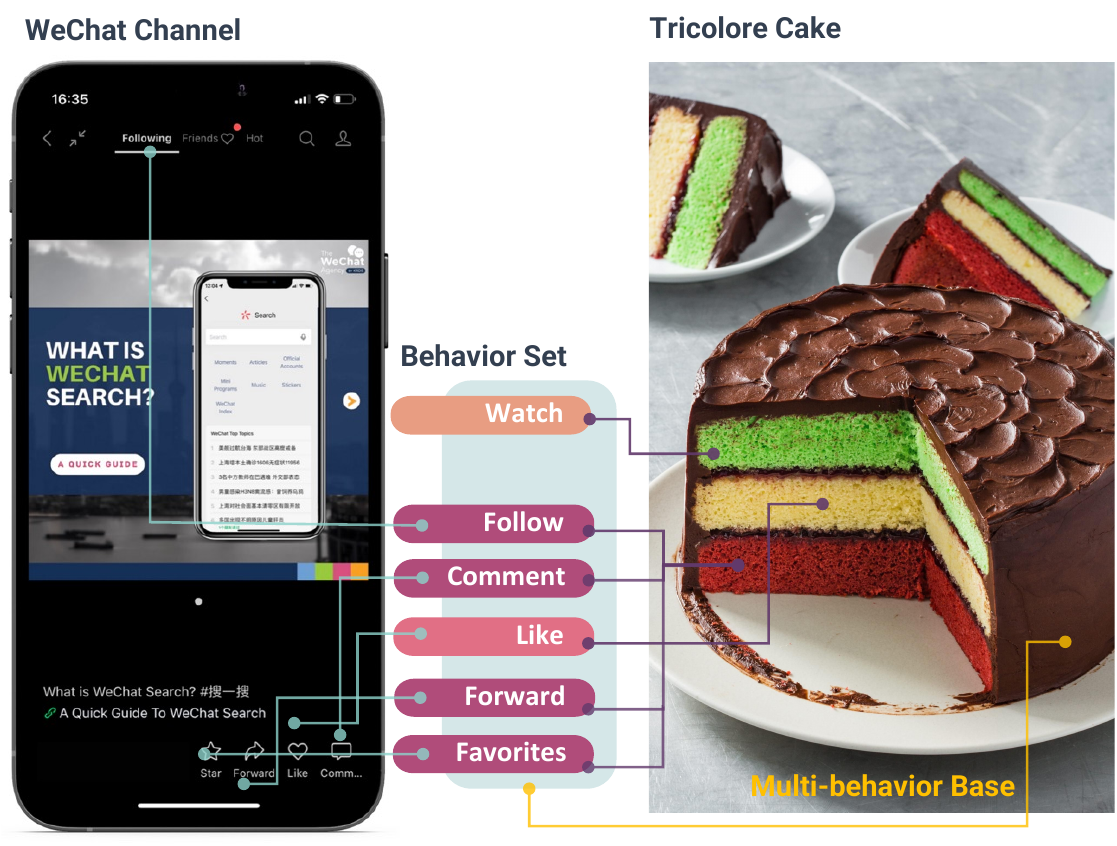}
    \caption{An illustration of multiple feedback types in WeChat Channels and the design inspiration of \textit{Tricolore}.}
    \label{fig:tricolore}
\end{figure}

One primary limitation of many existing MBRS is their tendency to manually \textbf{\textit{predetermine one primary behavior}} and treat other feedback types as auxiliary data to optimize recommendations on the target behavior~\cite{tang2016empirical,jin2020multi}. While these approaches generally outperform their single-behavior counterparts, they rigidly prioritize one behavior over others, which may not always align with practical needs. For instance, in e-commerce, while purchasing is typically considered the target behavior, sharing an item with friends can sometimes indicate stronger user preference. Similarly, on platforms like WeChat Channels\footnote{https://www.wechat.com}, where optimizing both viewing time and interaction rates is crucial, fixed targeting of one behavior may be inadequate (see Fig. \ref{fig:tricolore}). This limitation can undermine the system's ability to model diverse user interests, especially in early stages like candidate generation. Narrowly focusing on a single target behavior risks filtering out potentially appealing items reflected in other behaviors, complicating later stages of the recommendation process.

Another common limitation in current MBRS is the \textbf{\textit{insufficient detection of fine-grained interdependencies}} between different types of user behaviors. For instance, in e-commerce, behaviors like 'add-to-cart' and 'purchase' are typically positively correlated, while 'click' and 'hide' may exhibit negative correlations. Neglecting these associations in modeling can hinder the extraction of effective collaborative signals for user preference representation. Neglecting these associations in modeling can hinder the extraction of effective collaborative signals for user preference representation~\cite{xia2020multiplex}. Many existing MBRS tend to model individual behavior types separately and then integrate them simplistically for predicting target behaviors, often overlooking these interdependencies. While recent studies like chainRec~\cite{wan2018item} and NMTR~\cite{gao2019neural} have begun considering dependencies across feedback types, they often impose rigid sequential assumptions such as \textit{click}$\rightarrow$\textit{add to cart}$\rightarrow$\textit{purchase}~\cite{zhang2020multiplex}. We argue that this assumption is overly simplified and too rigid to be universally applicable in practical settings. For instance, it is common for users to either add an item to their favorites or share its link with friends. However, the sequence in which these behaviors occur can vary among different users~\cite{zhang2020multiplex}. This issue is further complicated in emerging domains like short video platforms, where behavioral sequences are less defined. Overall, user behavioral patterns across different types intertwine in complex ways that defy rigid predefined sequences. While some sequential patterns exist in specific contexts, models should dynamically learn these patterns rather than imposing fixed sequences manually~\cite{xia2021graph}.

In addition to identifying commonalities and connections, it is crucial to address the sharp \textbf{\textit{distinctions between individual behavior types}} in the development of MBRS. Each type of user behavior carries unique semantics that characterize diverse user preferences~\cite{xia2021multi,xia2021graph}. Moreover, these behaviors exhibit highly unbalanced sparsity levels, a factor often overlooked in current  MBRS approaches. Many existing multi-behavior recommendation algorithms rely on complex model structures with high computational demands, posing challenges for real-world industrial applications such as candidate generation tasks~\cite{yan2022cascading}. Hence, there is a pressing need for a multi-behavior recommendation framework capable of efficiently integrating multiple perspectives to capture both global commonalities and fine-grained distinctions among user behavior types simultaneously~\cite{wu2022multi,jin2020multi}.

To address the limitations of current MBRS techniques, we introduce \textit{Tricolore}, a novel framework designed for comprehensive and hierarchical depiction of user preferences in multi-behavior recommendation scenarios. Inspired by the Tricolore cake (Fig. \ref{fig:tricolore}), our model employs an elastic multi-bucket structure to simultaneously capture commonalities within each behavior class while preserving the unique characteristics of individual classes. A foundational layer detects associations between multiple behavior types, enhancing the learning process with supplementary information. We advocate for a lightweight and flexible multi-task model structure, enabling platforms to tailor prediction goals to specific recommendation needs. Given the varying degrees of sparsity across behavior types, our framework includes a behavior-wise multi-view fusion module to adaptively integrate global and local features. Additionally, we implement a popularity-balancing mechanism to mitigate the impact of popularity bias during negative sampling. Experimental evaluations on three public datasets validate \textit{Tricolore}'s effectiveness across diverse scenarios, encompassing short video and e-commerce recommendations.

Our primary contributions can be summarized as follows:
\begin{itemize}[leftmargin=*]
\item We propose a novel multi-behavior recommendation framework that shows superiorities in user preference modeling naturally by exploiting multiple types of historical user feedback as signals using more sophisticated multi-view fusion techniques. 

\item Leveraging a multi-bucket structure with the custom gate control mechanism and rich global information as the base for behavior-wise representation learning, \textit{Tricolore} is able to effectively capture profound associations between behavior types and deliver recommendations that match particular platform specifications.

\item We adopt a more meaningful negative sampling scheme to alleviate the popularity bias and strike a balance between accuracy and popularity in recommendations for item representation learning. 

\item The proposed \textit{Tricolore} model significantly improves the recommendation performance for the candidate generation task compared with competitive baseline methods on real-world datasets.
\end{itemize}

\section{Related Work}
In this section, we discuss existing literature on multi-behavior recommender systems (MBRS). 

The field of MBRS is relatively new, with one of the earliest studies conducted on LinkedIn products~\cite{tang2016empirical}. This study proposed three methods for incorporating multiple types of user feedback: training individual models in parallel, sequential training using prior feedback models, and joint training in a single optimization problem. The third method outperformed others, particularly when feedback types were correlated or when data for some behaviors was limited. This pioneering work has influenced subsequent studies and emphasized the importance of leveraging multiple user feedback types in industrial recommender systems.

MBRS studies are generally categorized into three approaches: \textbf{\textit{sampling-based}}, \textbf{\textit{loss-based}}, and \textbf{\textit{model-based}}. In the \textit{sampling-based} category, Loni et al.~\cite{loni2016bayesian} 
extended the vanilla Bayesian Personalized Ranking (BPR)~\cite{rendle2014improving} approach to distinguish the different strength levels of user preferences exhibited in various behavior types. Here, feedback types were ordered and endowed with weights according to their levels of importance, which would further influence how likely they were sampled in the training phase. Similarly, Ding et al.~\cite{ding2018improved} leveraged the information of behavior type of \textit{view} in e-commerce and proposed a view-enhanced sampler technique for classical BPR. From a loss-based point of view, Ding et al.~\cite{ding2018improving} also emphasized the importance of integrating \textit{view} data to advance recommendation performance using the same datasets. Compared to their earlier work~\cite{ding2018improved}, the authors improved the model by adding pair-wise ranking relations between \textit{purchase}, \textit{view}, and \textit{non-interacted actions} in the loss function instead of adopting point-wise matrix factorization methods. 

Another line of MBRS research targeted on the model modification. Among them, Liu et al.~\cite{liu2017personalized} utilized both explicit feedback (e.g. \textit{ratings}) and implicit feedback types (e.g. \textit{view}, \textit{click logs}) as input and employed explicit feedback to generate ordered partial pairs for training. Based on LSTM~\cite{hochreiter1997long} networks, Li et al.~\cite{li2018learning} designed a framework that was capable of learning short-term intention and long-term preferences of users through different behavior combinations for the next purchase item recommendation. Another work that adopted an RNN-based model and exploited sequential user behaviors for recommendation in e-commerce scenario~\cite{zhou2018micro} utilized feedback types of \textit{click, browsing, adding to cart, and dwell time}. Later, Zhou et al.~\cite{ zhou2019collaborative} proposed the Multi-Relational Memory Network based on an investigation of the strength and diversity levels of behavior types and adopted the attention mechanism to capture fine-grained user preferences across multi-behavior space. 

Some more recent investigations began to adopt graph neural network-based techniques for multi-behavior relational modeling. For instance, MBGCN~\cite{jin2020multi} was proposed to utilize the power of graph convolutional network (GCN) in learning complicated user-item and item-item connections for the multi-behavior recommendation. Similarly, CRGCN~\cite{yan2023cascading} utilizes a cascading residual graph convolutional network structure to learn user preferences by refining embeddings across multiple types of behaviors. MB-CGCN~\cite{cheng2023multi} employs cascading graph convolution networks to learn sequential dependencies in behaviors. MBSSL~\cite{xu2023multi} adopts a behavior-aware graph neural network with a self-attention mechanism to capture the multiplicity and dependencies of behaviors. Chen et al.~\cite{chen2021graph} proposed the GHCF model that focused on the use of heterogeneous high-hop structural information of user-item interactions in multiple types. Some other recent attempts include FeedRec~\cite{wu2022feedrec}, in which authors employed an attention network to distinguish user engagement levels on different feedback types for news recommendations. Another framework named MMCLR~\cite{wu2022multi} introduced contrastive learning (CL) in multi-behavior recommendations and designed three specific CL tasks to learn user representations from different views and behavior types.

In contrast to existing MBRS that either \textbf{\textit{prioritize a single behavior}} or rely on \textbf{\textit{rigid predefined sequences}} for behavior modeling, our proposed \textit{Tricolore} model introduces a flexible and adaptive framework capable of simultaneously capturing both global commonalities and fine-grained distinctions across multiple feedback types. One of the key advantages of \textit{Tricolore} is its ability to dynamically learn complex interdependencies between diverse behavior types without imposing overly simplistic or rigid assumptions, which is a limitation in many existing models. Furthermore, the inclusion of a popularity-balancing mechanism mitigates the impact of popularity bias, ensuring that the recommendation list not only prioritizes accuracy but also promotes diversity. These design choices enable \textit{Tricolore} to outperform state-of-the-art MBRS models, particularly in candidate generation tasks. The ability to tailor the multi-task structure to specific platform needs also enhances its practical applicability, making \textit{Tricolore} a versatile and robust solution for modern recommendation systems.

\section{Preliminaries}\label{preliminary}
\newcommand{\Ua}{\pazocal{U}}
\newcommand{\Va}{\pazocal{V}}
\newcommand{\Oa}{\pazocal{O}}
\newcommand{\Ya}{\pazocal{Y}}

In this section, we formulate our research problems and introduce the notations used throughout the paper.

\textit{Definition 1 (Candidate Generation):} Candidate generation in recommendation systems is crucial for selecting a personalized subset of items from a large pool, aligning with user preferences by predicting relevant items based on user behavior and item characteristics.

\textit{Definition 2 (Multi-Behavior Recommendation):} Let $\Ua$ denote the set of users and $\Va$ denote the set of items. The multi-behavior interaction tensor $\Ya \in \mathbb{R}^{|\Ua|\times |\Va|\times {K}}$ is defined to reflect multiple types of implicit user feedback, where each entry $y_{uv}^{k}$ in $\Ya$ records whether user $u \in \Ua$ has interacted with item $v \in \Va$ under the behavior type $k$. $K(K\geq2)$ is the number of user behavior types. Here given the behavior type $k$, $y_{uv}^{k}$ is set to 1 if the interaction between user $u$ and item $v$ is observed. Otherwise, it is assigned the value of 0. Different from most existing studies on multi-behavior recommendations where a target behavior type is pre-selected manually for optimization, the multi-behavior recommender system in this work adopts a flexible go-setting strategy and allows researchers or engineers to pursue tailor-made designs for different recommendation scenarios and the shift of purpose. More specifically, taking advantage of the multi-task learning structure in \textit{Tricolore}, one can either estimate the overall likelihood $\hat{y}_{uv}$ that user $u$ would enjoy a non-interacted item $v$ or her more refined preference on particular behaviors $\hat{y}_{uv}^{k}$ by fully utilizing the rich historical information across multiple types of user feedback and generate the Top-N item list for recommendation. 

\textit{Definition 3 (Behavior Class):} Online platforms often allow users to interact with items in various ways. \textit{Tricolore} suggests that platforms categorize these behavior types to address data sparsity and simplify the model structure. For each item $v$, we compute key indicators (e.g., 10s completion rate, purchase rate, like rate) for $K$ behaviors. Subsequently, across all items $\Va$, we calculate the correlation between pairs of these indicators and organize the $K$ behaviors into $C$ categories based on the correlation results $(C \le K)$. Behaviors with high positive correlations are grouped into the same category.

\section{Methodology}\label{method}
\newcommand{\Ha}{\pazocal{H}}
\newcommand{\Wa}{\pazocal{W}}
\newcommand{\La}{\pazocal{L}}
\newcommand{\Da}{\pazocal{D}}
\newcommand{\Qa}{\pazocal{Q}}

\begin{figure*}[h]
    \centering
    \includegraphics[width=0.9\textwidth]{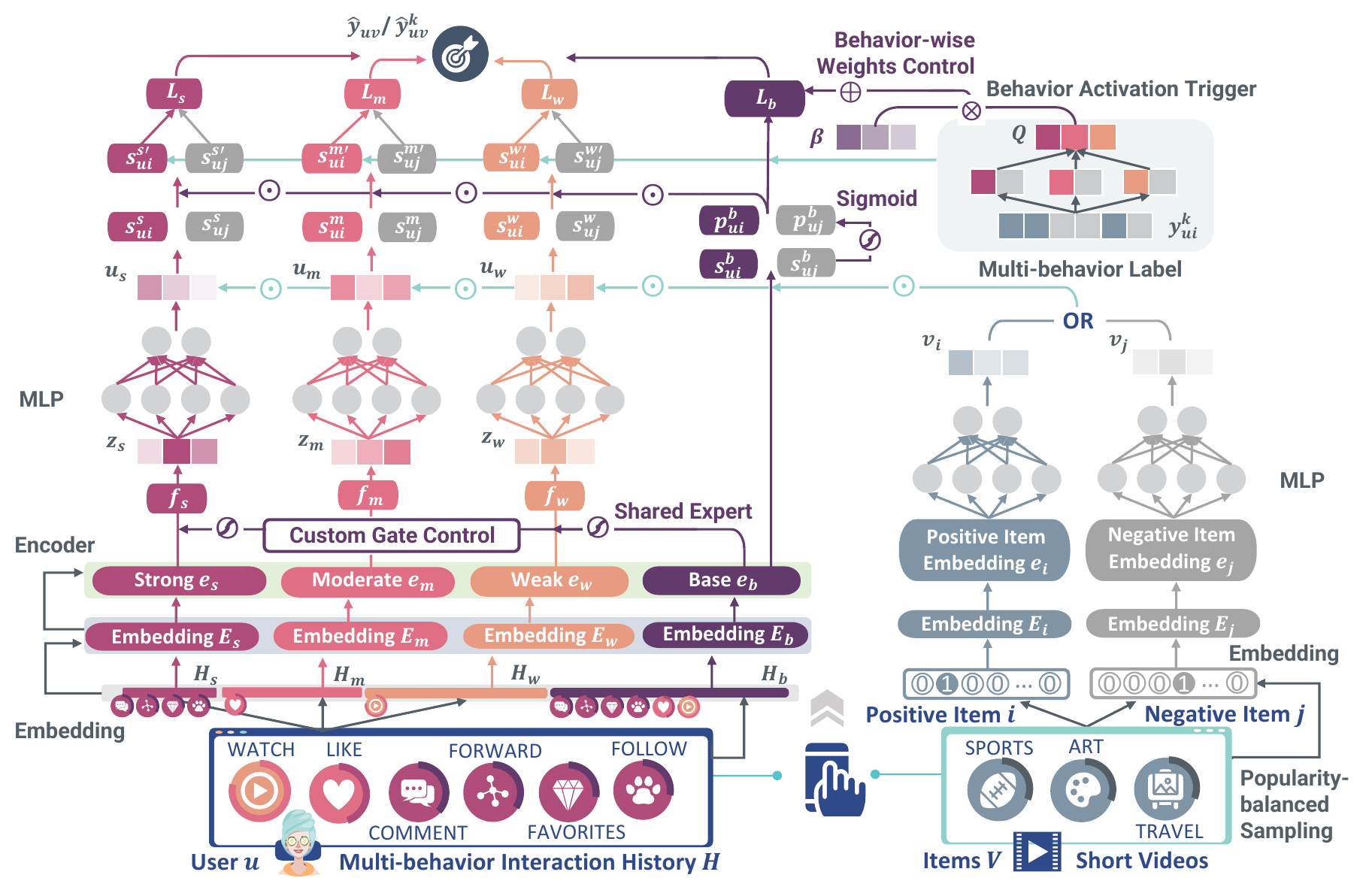}
    \caption{The architecture of the proposed \textit{Tricolore} framework.}
    \label{fig:architecture_model}
\end{figure*}

Next, we elaborate on the technical details of \textit{Tricolore}. \textbf{Overview.} Essentially, the \textit{Tricolore} framework is composed of four key modules: i) \textit{multi-behavior encoder} for initial user embedding learning on various behaviors; ii) \textit{behavior-wise multi-view fusion module} for global-information-enhanced user representation set formulation; iii) \textit{popularity-balanced item representation learning} for item embedding; iv) \textit{allied multi-task prediction module} for joint learning and the generation of multi-behavior recommendation lists. For a better illustration, we employ the short video recommendation scenario as an example to display the structural details of \textit{Tricolore} in graphical representation (Fig. \ref{fig:architecture_model}) and textual description below. 

\subsection{Multi-behavior Encoder}

Unlike most recommender systems that rely on a single type of behavior for user representation, or other multi-behavior approaches that merge feedback types into a single embedding, we introduce \textit{Tricolore}, which hierarchically constructs multi-vector user embeddings. These embeddings capture a more comprehensive view of user preferences across different behaviors.
We begin by categorizing $K$ user feedback types into $C$ groups based on similarities in user engagement and their impact on the platform. Using Pearson correlation analysis (as introduced in Section \ref{preliminary}), we classify behavior types. For instance, in the short video scenario, we select core indicators for each behavior, such as the \textit{10s completion rate} for \textit{watch}, the \textit{like rate} for \textit{like}, the \textit{comment rate} for \textit{comment}, the \textit{forward rate} for \textit{forward}, the \textit{favorite rate} for \textit{favorite}, and the \textit{follow rate} for \textit{follow}. In the e-commerce scenario, we use similar rates for \textit{purchase}, \textit{cart}, \textit{favorite}, and \textit{click}.

We then compute Pearson correlation coefficients and group behaviors with high correlations. In the short video scenario, \textit{watch} and \textit{like} show low correlations with other behaviors (below 0.1), while \textit{comment}, \textit{forward}, \textit{favorite}, and \textit{follow} have higher correlations (above 0.3). As a result, we classify behaviors into three categories: "strong," "moderate," and "weak," reflecting the user's affinity for items. For example, \textit{comment}, \textit{forward}, \textit{favorite}, and \textit{follow} are categorized as "strong," \textit{like} as "moderate," and \textit{watch} as "weak." In e-commerce, \textit{cart} and \textit{favorite} are grouped as "moderate" ( correlation 0.7630), \textit{purchase} is "strong," and \textit{click} is "weak."

Subsequently, we extract the interaction history of user $u$ with the "strong" behavior set, denoting it as $H_s$. We then create a one-hot embedding matrix, $E_s \in \mathbb{R}^{{K_s}\times d}$, which is further encoded to generate the initial embedding for "strong" behavior, denoted as $e_s \in \mathbb{R}^{d}$. Here, $K_s$ represents the number of behavior types in the "strong" category, and $d$ denotes the dimension of user embeddings. Formally:

\begin{equation}
\begin{split}
E_s &= g_{emb}(H_s\mid H_s \in \Ha),\\
e_s &= g_{enc}(E_s), 
\label{equation_1}
\end{split}
\end{equation}
where $g_{enc}(\cdot)$ is the encoder function for initial user representations learning that one can choose to adopt the average pooling, self-attention, or other attention mechanisms according to need here. $g_{emb}(\cdot)$ is the embedding operation that maps interaction sequences to vectors. $\Ha$ is the complete user muti-behavior interaction history.

In a similar way, we can respectively obtain the embeddings of user behavior types that show moderate and weak liking, which are $e_m \in \mathbb{R}^{d}$ and $e_w \in \mathbb{R}^{d}$. Apart from these representations of user, we also learn a base embedding $e_b \in \mathbb{R}^{d}$ for the capture of the global preference of the user, which takes $K$ types of user behaviors into consideration and serves as a significant supplement for individual behavior classes in the \textit{behavior-wise multi-view fusion module} next.


\subsection{Behavior-wise Multi-view Fusion}\label{BW_multi-view}

The motivation behind \textit{Tricolore}'s multi-bucket model and multi-view fusion strategy is to address sparsity issues in recommender systems while capturing both local preferences at each behavior level and global user preferences.

Most existing systems focus on a single primary behavior, like \textit{click} in advertising or \textit{purchase} in e-commerce. Although these behaviors are tightly linked to platform profit and user intent, they often suffer from data sparsity. While multi-behavior models help alleviate this, further improvements are needed to better model user preferences across behaviors. To this end, we propose the \textit{behavior-wise multi-view fusion module}, which uses a custom gate control mechanism to adjust user embeddings by allowing behaviors to draw on shared expert knowledge $e_b$. We introduce learnable weight vectors $\Wa = \{W_s, W_m, W_w \}$, which let behaviors decide how much information to borrow from other behaviors. For instance, the \textit{strong} behavior class is modeled as:

\begin{equation}
\begin{split}
f_s &= \sigma (e_s W_s^{T}),\\
z_s &= f_s e_b + (1-f_s) e_s, 
\label{equation_2}
\end{split}
\end{equation}
where $\sigma(\cdot)$ denotes the sigmoid function. $W_s \in \mathbb{R}^{d}$ is the weight vector of \textit{strong} behaviors. $f_s$ is a learnable parameter that determines the extent to which information is incorporated from the \textit{base} knowledge into the representation of \textit{strong} behaviors. $z_s \in \mathbb{R}^{d}$ is the revised \textit{strong} user embedding. 

We then utilize $z_s$ as the input of a multilayer perceptron (MLP) to obtain the final \textit{strong} user embedding $u_s \in \mathbb{R}^{d}$ as:
\begin{equation}
u_s = MLP (z_s).
\label{equation_3}
\end{equation}
In a similar vein, \textit{Tricolore} can generate the final user embeddings of $u_m$ and $u_w$ for \textit{moderate} and \textit{weak} behavior types, respectively.

This multi-view fusion approach enables us to learn user representations that retain independent components at each behavior level, while sharing useful knowledge across behaviors. A key feature is the use of learnable parameters in the gate control for each behavior type.

For \textit{strong} behaviors (e.g., \textit{comment}, \textit{forward}, \textit{favorite}, \textit{follow} in short video), the contribution from the base embedding $e_b$ is expected to be larger due to sparsity. In contrast, \textit{watch} behavior, though weaker, is still informative. Interestingly, stronger behaviors tend to correlate positively with \textit{watch}, as users often engage in stronger interactions after watching for more than 10 seconds. Therefore, the multi-view fusion technique improves the quality of \textit{weak} embeddings more by purifying the information rather than supplementing it. This is discussed in detail in Section \ref{associations}.

In special cases, such as cold-start users with limited interaction history across any behavior category, the shared base embedding $e_b$ becomes even more crucial in shaping user representations. The varying parameters learned across different behavior types in this module, along with the performance of \textit{Tricolore} in cold-start settings, are discussed in Section \ref{cold}.

\subsection{Popularity-balanced Item Representation}\label{popularity_sampling}

For item embedding learning in \textit{Tricolore}, the identities of a pair of positive and negative items, $i$ and $j$, are initially represented as two binary sparse vectors $E_i$ and $E_j$ via one-hot encoding. Here positive item $i\in V_{u}^{+}$ is a video that user $u$ has interacted with on more than one behavior type, while negative item $j\in V_{u}^{-}$ means that no interaction between user-item pair $\left \langle u,j \right \rangle$ is observed. Then they are projected to low-dimensional dense vectors to generate item embeddings, denoted as $e_i \in \mathbb{R}^{d}$ and $e_j \in \mathbb{R}^{d}$. After that, MLP is implemented to learn the final item embeddings $v_i$ and $v_j$ through:
\begin{equation}
v_i = MLP (e_i);\quad v_j = MLP (e_j).
\label{equation_4}
\end{equation}
The dimension of the generated vectors $v_i$ and $v_j$ is also $d$.

Considering the ubiquitous popularity bias problem that haunts many recommender systems~\cite{zhang2021causal}, the proposed \textit{Tricolore} framework adopts a popularity-balanced negative sampling technique inspired by that in~\cite{mikolov2013distributed} to penalize the sampling probability given to items according to their historical popularity. Mathematically, the probability for item $v_j$ to be selected as a negative sample $P_n(v_j)$ can be computed by the formula:
\begin{equation}
P_n(v_j)=\frac{c(v_j)^{\gamma}}{\sum_{j}^{}(c(v_j)^{\tau})},\forall j = 1,2,\cdots,|V_{u}^{-}|,
\label{equation_5}
\end{equation}
where $c(v_j)$ is the frequency of item $v_j$ that emerged in the historical interactions of all users $\Ua$, which reflects the overall popularity of the item. $\gamma$ and $\tau$ denote smoothness powers.



The core idea behind this approach is to mitigate item popularity bias by increasing the negative sampling probability for more popular items. This encourages a more balanced distribution of negative samples and enhances the diversity of recommendation lists. The goal is to ensure that both popular and niche items are fairly represented during model training, preventing overfitting to frequently interacted items. This balanced treatment improves the model's generalization capability and promotes greater diversity in recommendations.

In recommender systems, particularly in two-tower models like \textit{Tricolore}, this strategy helps address the challenge of imbalanced exposure between popular and less popular items. Our method specifically targets the imbalance issue in multi-behavior scenarios, where users engage with a wide range of behaviors. This is especially important, as popularity bias can blur behavior distinctions, and our approach ensures fairness in how items are treated across different behavior types.

The trade-off between recommendation accuracy and item popularity bias, as well as how to properly adjust the smoothness powers to strike the right balance, will be discussed in Subsection \ref{trade-off}.

\subsection{Allied Multi-task Prediction}

Thus far, we have obtained multi-behavior user preference representations and embeddings for both positive and negative items. It is fairly straightforward to calculate the prediction scores by dot product for each user-item pair of embeddings. For instance, the scores for base user embedding and items can be achieved via:
\begin{equation}
s_{ui}^{b} = e_{b} \cdot v_{i}; \quad s_{uj}^{b} = e_{b} \cdot v_{j}.
\label{equation_6}
\end{equation}
We can also calculate the initial scores for \textit{strong} user behaviors as:
\begin{equation}
s_{ui}^{s} = u_{s} \cdot v_{i}; \quad s_{uj}^{s} = u_{s} \cdot v_{j}.
\label{equation_7}
\end{equation}
Similarly, the relevant initial scores for moderate and weak user behaviors $(s_{ui}^{m},s_{uj}^{m})$ and $(s_{ui}^{w},s_{uj}^{w})$ can be also obtained, respectively.

However, considering the underlying sequential patterns may exist among user interactive behaviors, i.e., $click\rightarrow cart \rightarrow purchase$ or $watch\rightarrow like/forward/follow/comment$, we suggest taking one step further and propose an \textit{allied multi-task learning} scheme. The key assumption here is that user behaviors do not exist independently but have mutual influences on each other. Only when a user shows basic interest in an item, he/she would create the following behaviors, such as following the author or forwarding the item to a friend. Instead of determining the sequential pattern of influence manually as most current multi-behavior studies do, we advocate a more elastic plan. Specifically, we convert the scores of base embedding into probabilities first by employing the \textit{sigmoid} function as:
\begin{equation}
p_{ui}^{b} = \sigma(s_{ui}^{b}); \quad p_{uj}^{b} = \sigma(s_{uj}^{b}).
\label{equation_8}
\end{equation}
We then tune the initial prediction scores of \textit{strong} behaviors to get its final version  $({s_{ui}^{s}}',{s_{uj}^{s}}')$ by:
\begin{equation}
{s_{ui}^{s}}'= p_{ui}^{b} {s_{ui}^{s}}; \quad {s_{uj}^{s}}'= p_{uj}^{b} {s_{uj}^{s}}.
\label{equation_9}
\end{equation}
Similarly, we calculate the prediction scores for moderate and weak user feedback types, $({s_{ui}^{m}}',{s_{uj}^{m}}')$ and $({s_{ui}^{w}}',{s_{uj}^{w}}')$, which signal the likelihood that user $u$ would be interested in an item on particular behavior classes. The higher the score, the more likely it would be. 

\subsection{Optimization}

We leverage the pair-wise ranking loss for optimization in \textit{Tricolore}. Instead of treating unobserved entries as negative feedback in point-wise loss \cite{hu2008collaborative}, pair-wise learning \cite{rendle2009bpr} focuses on the relative positions of each observed-unobserved pair of items that observed entries are expected to rank higher than their counterparts. Mathematically, the objective function for each behavior level can be defined as:  
\begin{equation}
\begin{split}
\La_s &= \sum_{(u,i,j)\in \Da_s}^{} max(0, {s_{uj}^{s}}'-{s_{ui}^{s}}'+\lambda ^s),\\
\La_m &= \sum_{(u,i,j)\in \Da_m}^{} max(0, {s_{uj}^{m}}'-{s_{ui}^{m}}'+\lambda ^m),\\
\La_w &= \sum_{(u,i,j)\in \Da_w}^{} max(0, {s_{uj}^{w}}'-{s_{ui}^{w}}'+\lambda ^w),
\label{equation_10}
\end{split}
\end{equation}
here $\Da_s$, $\Da_m$, and $\Da_w$ denote the sub-datasets under \textit{strong}, \textit{moderate}, and \textit{weak} behavior levels. $\lambda ^{s}$, $\lambda ^{m}$, and $\lambda ^{w}$ represent the safety margin sizes within the pairwise hinge loss function, which serve to separate negative items from positive items within each behavior class.

For the base loss function design in \textit{Tricolore}, we consider both cross-class behavior dependencies and class-specific semantics by:
\begin{equation}
\La_b=\sum_{(u,i,j)\in \Da} max(0, {s_{uj}^{b}}'-{s_{ui}^{b}}'+\lambda ^b)\left (\beta _sQ_s +\beta _mQ_m+\beta _wQ_w \right),
\label{equation_11}
\end{equation}
where $\Da$ represents the whole dataset on various behavior types. $\lambda ^{b}$ is the margin separating positive and negative items in \textit{base} view. $\beta _s$, $\beta _m$, and $\beta _w$ are \textit{behavior-wise control weights} to determine how much each behavior level contributes to the \textit{base} loss $\La_b$. $\Qa = \left \{Q_s, Q_m, Q_w \right \}$ denotes \textit{behavior activation trigger} that indicates interactions are observed on which behavior types for each positive sample $\left \langle u,i \right \rangle$. Here each element in $\Qa$ is a binary indicator vector that contains 1 for the observed behavior class and 0 otherwise. $\Qa$ determines which branch(es) of behavior types would be activated for the individual level loss calculation and the \textit{base} loss $\La_b$.

The overall objective function for \textit{Tricolore} is a weighted sum of all the contributions of each related individual objective above as:
\begin{equation}
\La = \La_b + \alpha \La_s + \varepsilon \La_m + \zeta \La_w + \mu \left \|\theta  \right \|_{2}^{2},
\label{equation_12}
\end{equation}
here $\alpha$, $\varepsilon$, and $\zeta$ are weighting parameters specified to influence to what degree each behavior level-specific effect should be taken into account during the optimization. $\theta$ is the model parameter set; and $\mu$ is a parameter that controls the importance of the last term, where we apply regularization to prevent overfitting by using the dropout strategy and adding $L_2$-norm terms. 

It is also worth noting that \textit{Tricolore} adopts a flexible multi-task framework that allows platforms to choose to predict the overall likelihood that a user would enjoy an item $\hat{y}_{uv}$ or his/her interest on a type-specific behavior $\hat{y}^k_{uv}$. Besides, by injecting discriminative $\lambda$, we can encode the \textit{strength} deviation corresponding to different behavior levels and views. Here a larger setting of margin would push the positive and negative samples farther away from each other, suggesting that this behavior class or view is more reliable in depicting user preferences.

\section{Experiments}\label{experiment}
\subsection{Experimental Settings}
\subsubsection{Datasets}
The experiments are conducted using publicly available datasets from a short-video application and two e-commerce platforms, which are described below.

\noindent \textbf{WeChat Channels\footnote{https://algo.weixin.qq.com/2021/problem-description}.} This is a short-video dataset released by WeChat Big Data Challenge 2021. It contains six types of user-video interactions on the WeChat Channels application. 

\noindent \textbf{Tmall\footnote{https://tianchi.aliyun.com/dataset/dataDetail?dataId=649}.} This is an open dataset from Tmall\footnote{https://www.tmall.com}, one of the largest e-commerce platforms in China. It contains four typical types of user behavior in e-commerce scenarios, including \textit{click, favorite, add-to-cart}, and \textit{purchase}.

\noindent \textbf{CIKM\footnote{https://tianchi.aliyun.com/competition/entrance/231721/introduction}.} It is offered by the CIKM 2019 EComm AI Challenge and includes the same user behaviors as the Tmall dataset.

We weed out users with less than ten interactions for the CIKM data and three interactions for WeChat Channels and Tmall datasets, respectively. To counter the imbalance between the rare and frequent user feedback types, the datasets are intentionally sampled so that each of them is well represented in the training phase. The statistics of the processed datasets are summarized in Table \ref{tab:data}. Here we color the cells for behavior types according to which bucket they belong to in \textit{Tricolore} for the experiments. Although as a multi-task learning framework, \textit{Tricolore} does not require us to specify a single target behavior to optimize, we use '*' to denote a primary behavior type generally employed by traditional single-behavior recommender systems in each scenario for further comparative analyses.

\begin{table}[h]
	\caption{Statistics of the datasets. '*' denotes the primary behavior type in baseline models if applicable.}
	\label{tab:data}
	\centering
    \renewcommand\arraystretch{1.3}
    \setlength{\tabcolsep}{2.1mm}
	\renewcommand\arraystretch{1.2}
	\footnotesize
	\begin{tabular}{c|c||c|c|c|c}
		\hline 
		\bf Dataset&\bf Behavior&\bf \#User&\bf \#Item&\bf \#Inter&\bf Density\\
		\hline
		\hline  
		\multirow{6}*{\bf Channels}&\cellcolor{Wcolor}{\color{white}Watch$^*$} &  218,919& 44,846 & 2,157,358 & 0.022\% \\
		\cline{2-6}
		&\cellcolor{Mcolor}{\color{white}Like} & 179,270& 24,176&  897,019& 0.020\% \\
		\cline{2-6}
		&\cellcolor{Scolor}{\color{white}Comment} & 13,649 & 1,869 & 35,156 & 0.138\% \\
		\cline{2-6}
		&\cellcolor{Scolor}{\color{white}Forward} & 81,792 & 9,059 & 281,547 &0.038\% \\
		\cline{2-6}
		&\cellcolor{Scolor}{\color{white}Favorite} & 25,507 & 3,953 & 90,989 & 0.090\% \\
		\cline{2-6}
		&\cellcolor{Scolor}{\color{white}Follow} & 26,473 & 3,312 & 55,513 &0.063\% \\
		\hline 
		\multirow{4}*{\bf Tmall}& \cellcolor{Scolor}{\color{white}Purchase$^*$} &4,760&  3,037& 6,237 & 0.043\% \\
		\cline{2-6}
		&\cellcolor{Mcolor}{\color{white}Cart} & 1,958 &2,353  & 3,588 & 0.078\% \\
		\cline{2-6}
		&\cellcolor{Mcolor}{\color{white}Favorite} & 980 & 1,502 &  1,993&0.135\% \\
		\cline{2-6}
		&\cellcolor{Wcolor}{\color{white}Click} & 4,675 & 66,927&  76,536&0.245\% \\
		\hline 
		\multirow{4}*{\bf CIKM}&\cellcolor{Scolor}{\color{white}Purchase$^*$} & 13,673 & 10,629 & 17,051 & 0.012\% \\
		\cline{2-6}
		&\cellcolor{Mcolor}{\color{white}Cart}& 3,208 & 3,544 & 4,260 & 0.037\% \\
		\cline{2-6}
		&\cellcolor{Mcolor}{\color{white}Favorite} & 1,298 & 1,624 & 1,765 & 0.083\% \\
		\cline{2-6}
		&\cellcolor{Wcolor}{\color{white}Click} & 13,325 & 26,216 & 59,659 & 0.017\% \\
		\hline 
	\end{tabular}
\end{table}

\subsubsection{Baselines}
To examine the effectiveness of the proposed \textit{Tricolore}, a series of state-of-the-art baselines are employed for comparison. It should be mentioned that for multi-behavior recommendations, only models applicable to candidate generation tasks are selected. Besides, to guarantee the fairness in comparison and dispel doubts on the data size inequality caused by involving more feedback types in MBRS, we treat all behavior types as primary feedback in the single-behavior recommender systems. These baseline models include:

\noindent \textit{Classic Single-behavior Recommendation Algorithms.}
\begin{itemize}[leftmargin=*]
\item \textbf{MF-BPR}~\cite{rendle2009bpr}: Bayesian Personalized Ranking is a classic method for item recommendation from implicit feedback, which is directly optimized for ranking.

\item \textbf{DSSM}~\cite{huang2013learning}: DSSM is an effective two-tower model for large-scale industrial recommender systems which makes predictions by matching the query and documents. 

\item \textbf{NCF}~\cite{he2017neural}: This is a typical recommendation algorithm that augments collaborative filtering with deep neural networks. 
\end{itemize}

\noindent \textit{Multi-behavior Recommendation Algorithms.}

\begin{itemize}[leftmargin=*]
\item \textbf{MC-BPR}~\cite{loni2016bayesian}: It is a sampling-based MBRS that samples positive and negative instants from multiple relations. 

\item \textbf{FeedRec}~\cite{wu2022feedrec}: FeedRec is a recent MBRS model proposed for news recommendation based on an attention network. 

\item \textbf{MBGCN}~\cite{jin2020multi}: It is a graph model with the ability to capture user-item and item-item level multi-behavior information.

\item \textbf{MMCLR}~\cite{wu2022multi}: MMCLR aims to predict the target behavior by the construction of contrastive learning tasks and the fusion strategy of the sequence model and graph model. 

\item \textbf{CRGCN}~\cite{yan2023cascading}: CRGCN utilizes a cascading residual graph convolutional network structure to learn user preferences by refining embeddings across multiple types of behaviors.

\item \textbf{MB-CGCN}~\cite{cheng2023multi}: It employs cascading graph convolution networks to learn sequential dependencies in behaviors.

\item \textbf{MBSSL}~\cite{xu2023multi}: It adopts a behavior-aware graph neural network that incorporates a self-attention mechanism to capture the multiplicity and dependencies of behaviors.
\end{itemize}

\noindent \textit{Multi-task Learning Algorithms.}

\begin{itemize}[leftmargin=*]
\item \textbf{MMoE}~\cite{ma2018modeling}: It adapts MoE structure to multi-task learning, allowing expert submodels to be shared across tasks within a gating network optimized for each individual task. 

\item \textbf{ESMM}~\cite{ma2018entire}: ESMM leverages sequential patterns in user actions and employs a transfer learning strategy to mitigate sample selection bias and address data sparsity issues.

\item \textbf{PLE}~\cite{tang2020progressive}: It is a multi-task learning approach that employs a progressive routing mechanism to differentiate the semantic knowledge of shared and task-specific components.

\end{itemize}

\subsubsection{Evaluation Metrics}
The performance of the models is evaluated mainly by two widely-used ranking metrics, hit ratio (\textit{HR@K}) and normalized discounted cumulative gain (\textit{NDCG@K}), where \textit{K} is set to $\left\{5, 10\right\}$. Besides, since the popularity-balanced strategy is adopted in the negative sampling phase, apart from metrics in accuracy, we also pay close attention to the model's performance in popularity bias. Here, we employ the average popularity of the top-10 item recommendation lists (ARP metric in \cite{abdollahpouri2019managing}) for evaluation. The positive item is compared with 99 negative samples in the test stage to evaluate the ranking performance of the models.

\subsubsection{Parameters Settings}

We implement \textit{Tricolore} using Tensorflow 2.0.0 and fine-tune the hyperparameters by grid-search according to its performance on the validation set. We search the sequence of behavior sample lengths in $\left\{1,3,5,10\right\}$, embedding size in $\left\{16,32,64,128\right\}$, and dropout rate \cite{srivastava2014dropout} in $\left\{0,0.1,0.2,0.3,0.4,0.5\right\}$. The model’s embedding size is set to 32; the dropout rate is 0.1. The margin of $\lambda$ employed in pair-wise ranking is explored among $\left\{0.05,0.1,0.3\right\}$ and set to 0.1. The weighting parameters of $\alpha$, $\varepsilon$, and $\zeta$ in the overall loss are set to $(0.1,0.1,0.1)$. We optimize all the baseline models as well as \textit{Tricolore} by the Adam optimizer \cite{kingma2014adam} and search the learning rate in $\left\{0.1,0.01,0.005,0.001,0.0001\right\}$. All models’ batch size is set to 256 and other parameters are followed by default settings according to the respective papers.

\subsection{Performance Comparison}

Table \ref{tab:overall_exp} displays the performance of \textit{Tricolore} and the baseline models across datasets. To ensure reliability, we conduct five tests for each method and average the outcomes for the final results. The analysis yields the following observations:

\begin{table*}[]
\caption{Experimental results are presented for the three datasets, with the best performance highlighted in boldface and the best baseline model marked with '*'. Reported mean and standard deviation values are based on the results of 5 random runs.}
	\label{tab:overall_exp}
	\centering
	\setlength{\abovecaptionskip}{0.09 cm}
	\setlength{\belowcaptionskip}{0 cm}
	\renewcommand\arraystretch{1.3}
    \setlength{\tabcolsep}{0.72 mm}
	\footnotesize
	\begin{tabular}{c||c|c|c|c||c|c|c|c||c|c|c|c}
		\hline 
		\multirow{2}*{\bf Methods}&\multicolumn{4}{c||}{\bf WeChat Channels}&\multicolumn{4}{c||}{\bf Tmall}&\multicolumn{4}{c}{\bf CIKM}\\[0.75 ex]
		\cline {2-13}
		&\bf HR@5&\bf HR@10&\bf NDCG@5&\bf NDCG@10&\bf HR@5&\bf HR@10&\bf NDCG@5&\bf NDCG@10&\bf HR@5&\bf HR@10&\bf NDCG@5&\bf NDCG@10\\
		\hline 
		\hline 
		\bf MF-BPR&  0.3009&0.4389 & 0.2027& 0.2470 &0.4498  &0.4830  &0.4069 &0.4177  & 0.2425 & 0.3526&0.1428& 0.1843 \\
		\hline 
        \bf DSSM&  0.2479&0.3834 & 0.1619& 0.2057&  0.3345 &  0.4045& 0.2629 &  0.2854 &0.2440 &0.3520&0.1466 &0.1815 \\
		\hline 
		\bf NCF&  0.2594&  0.4044& 0.1691 &0.2156 &0.3663  &0.4186  & 0.3061 &0.3231  & 0.2637& 0.3457 & 0.1867&0.2130\\
		\hline 
		\bf MC-BPR& 0.3216&0.4618 &  0.2173 &0.2624& 0.3720 &  0.4306&  0.3123 &0.3310  &0.2701& 0.3479&0.1957 &0.2205\\
		\hline 
		\bf FeedRec&  0.2269& 0.3510 & 0.1456& 0.1855& 0.4470&0.4886 &  0.4040& 0.4182 & 0.2651& 0.3620&0.1821  &0.2133\\
		\hline 
		\bf MBGCN&0.2294 &0.3726  &0.1440  & 0.1898 &0.2258  & 0.3512 & 0.1471 & 0.1871 & 0.1910 & 0.2438 & 0.1420 & 0.1591\\
		\hline 
		\bf MMCLR&0.2106  & 0.3264 &0.1413 &0.1782  & 0.3330 &0.4214  & 0.2395 & 0.2679 & 0.2704 &0.3475  &0.1932& 0.2179  \\
  		\hline 
      	\bf CRGCN&0.3060&0.4502&0.2016&0.2479&0.5078$^*$&0.5255$^*$&0.4825$^*$&0.4881$^*$&0.2924&0.3638&0.2242$^*$&0.2470$^*$\\
		\hline 
		\bf MB-CGCN&0.2751&0.5424&0.1477&0.2340&0.4965&0.5120&0.4674&0.4723&0.2921&0.3673$^*$&0.2146&0.2386\\
		\hline 
		\bf MBSSL&0.4097$^*$&0.5680$^*$&0.2277$^*$&0.3062$^*$&0.4986&0.5099&0.4634&0.4670&0.2883&0.3638&0.2142&0.2384\\
		\hline
		\bf MMoE&0.3083 &0.4530 & 0.2082& 0.2549& 0.4052 &0.4589 & 0.3540&0.3719 & 0.2730&0.3411&0.1790&0.2009 \\
		\hline 
		\bf ESMM& 0.3187& 0.4682& 0.2084&0.2566 & 0.2765 &0.3338 &0.2232 &0.2417 & 0.3035$^*$ & 0.3608 & 0.2032 & 0.2217\\
		\hline 
		\bf PLE& 0.3171 & 0.4571 & 0.2114 & 0.2547 & 0.4490 & 0.4873 &0.4069 & 0.4191 & 0.2745 & 0.3373 & 0.1797 & 0.1999\\
		\hline 
        \bf Ours & \textbf{0.4564} & \textbf{0.6134} & \textbf{0.3262} & \textbf{0.3769} & \textbf{0.5677} & \textbf{0.5890} & \textbf{0.5108} & \textbf{0.5290} & \textbf{0.4288} & \textbf{0.4797} & \textbf{0.2943} & \textbf{0.3186}\\
            \hline
        \textbf{Impr.} & +11.40\% & +7.99\% & +43.26\% & +23.09\% & +11.80\% & +12.08\% & +5.87\% & +8.38\% & +41.29\% & +30.60\% & +31.27\% & +28.99\%\\
            \hline 
	\end{tabular}
	\captionsetup{margin=0cm}
\end{table*}

\subsubsection{Performance of \textit{Tricolore}} 
As depicted in Table \ref{tab:overall_exp}, \textit{Tricolore} demonstrates state-of-the-art performance across all three datasets, outperforming baseline models in four evaluation metrics. Particularly noteworthy is its superiority over the best baseline model by over 40\% in \textit{NDCG@5} on WeChat Channels and in \textit{HR@5} on CIKM.

\subsubsection{Single-behavior Vs. Multi-behavior Methods} When comparing the performance of MBRS baseline models with their single-behavior counterparts, it is evident that the MBRS class as a whole offers superior results. Specifically, MBSSL stands out as the best baseline model across all evaluation metrics on WeChat Channels, while CRGCN performs exceptionally well in e-commerce scenarios on Tmall and CIKM.

In a nutshell, the overall experimental results demonstrate the effectiveness and generalization capabilities of \textit{Tricolore} in multi-behavior recommendation tasks across various scenarios. This can probably be attributed to our fine-grained multi-vector learning strategy adopted in user representation and behavior-wise multi-view fusion mechanism in \textit{Tricolore} framework, which will be discussed more in the following ablation study.

\subsection{Ablation Study}

We further conduct the ablation study over key components of \textit{Tricolore} to better understand the effects of each individual module. More specifically, we introduce the following four variants of the model:

\begin{itemize}[leftmargin=*]
\item \textbf{w/o MVF}: The behavior-wise multi-view fusion module is removed in the stage of user preference modeling.

\item \textbf{w/o MVL}: It removes the multi-vector learning strategy, treats all user-item interactions as input regardless of behavior types, and learns a single vector for user representation.

\item \textbf{w/o MTL}: This variant changes the multi-task learning framework of \textit{Tricolore} to a single-task structure. 

\item \textbf{w/o AMT}: The allied learning scheme in the final prediction phase is removed that base probabilities are not utilized here. 

\item \textbf{w/o CAT}: This is a variant without categorization that employs individual embeddings per behavior for prediction.

\end{itemize}

To see the evaluation results in Table \ref{tab:ablation_exp}, we can draw the conclusion that all the key modules contribute to the overall performance of \textit{Tricolore}. Among them, the multi-task learning strategy plays the most significant role that the performance would drop by around 20\% in all the metrics without it. In addition, there are no evident differences among the contributions from the other four modules that a relative reduction of less than 10\% is observed in each of the metrics.

\begin{table}[]
	\caption{Ablation study on WeChat Channels dataset.}
	\label{tab:ablation_exp}
	\centering
	\setlength{\abovecaptionskip}{0 cm}
	\setlength{\belowcaptionskip}{0 cm}
	\setlength{\tabcolsep}{2.4mm}
	\renewcommand\arraystretch{1.3}
	\footnotesize
	\begin{tabular}{c||c|c|c|c}
		\hline 
		&\bf HR@5&\bf HR@10&\bf NDCG@5&\bf NDCG@10\\
		\hline 
        \hline 
		\textbf{w/o MVF} &0.3087&0.4614 &0.2037 &0.2539\\
		\hline 
        \textbf{w/o MVL} &0.3034 &0.4610&0.1985 &0.2493 \\
		\hline 
		\textbf{w/o MTL} &0.2479&0.3834 & 0.1619&0.2057\\
		\hline 
		\textbf{w/o AMT} &0.3057 &0.4427&0.2062 &0.2501\\
		\hline 
  		\textbf{w/o CAT} &0.2882 &0.4420 &0.1918&0.2412\\
		\hline 
		\textbf{Tricolore} &\textbf{0.3230} & \textbf{0.4737} &\textbf{0.2189}& \textbf{0.2674}\\
		\hline 
	\end{tabular}
	\captionsetup{margin=0cm}
\end{table}

\subsection{Associations Between Behaviors}\label{associations}

Next, we study the underlying associations between each individual behavior category and other behavior buckets as a whole, which correspond to local and global views, respectively. \textit{Tricolore} enables us to do such analyses via introducing the custom gate control mechanism in the \textit{Behavior-wise Multi-view Fusion} module. 

\subsubsection{Custom Gate Control Parameter $f$}
\begin{figure}[t]
    \centering
    \includegraphics[width=0.35\textwidth]{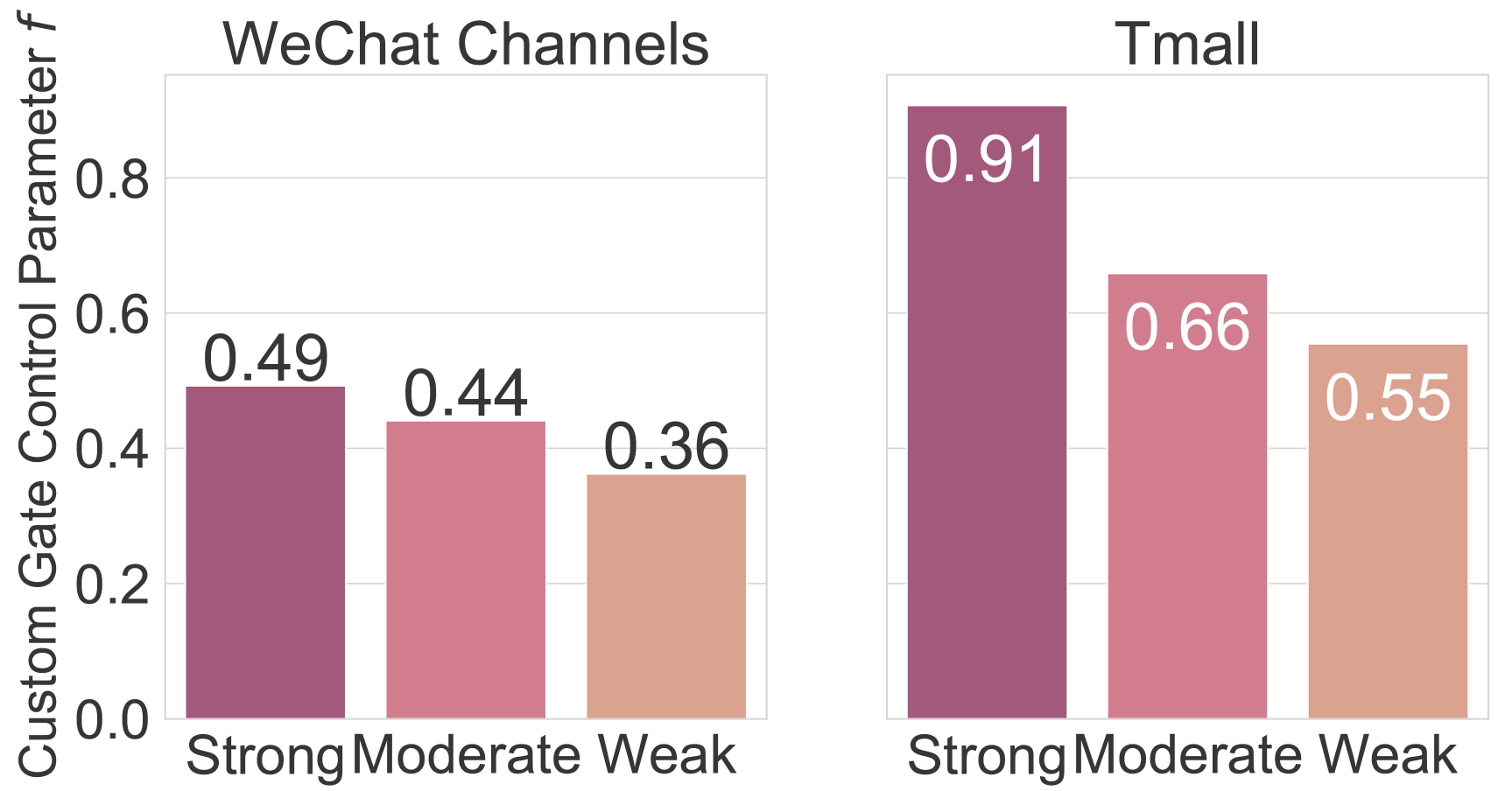}
    \caption{Study of custom gate control parameters $f$ on WeChat and Tmall.}
    \label{fig:Gate_Control_f}
\end{figure}

As introduced in Subsection \ref{BW_multi-view}, the base embedding $e_b$ that conveys information from all types of historical interactions of the user serves as a shared expert to assist in the representation learning on each behavior category according to their needs. With regard to how much help the behavior types in each branch prefer to get from the \textit{base} expert, we represent it by a learnable gate control parameter $f$. Here a larger value indicates more help is needed (see Eq.\ref{equation_2}). In Fig. \ref{fig:Gate_Control_f}, we compare the custom gate control parameters of \textit{strong}, \textit{moderate}, and \textit{weak} behaviors, $\left \{f_s,f_m,f_w \right \}$, learnt by \textit{Tricolore} on the datasets of WeChat Channels and Tmall. As can be seen from the figure, behavior types falling into the \textit{strong} group have the highest values of $f$ on both datasets. This phenomenon is consistent with our expectations that strong user behaviors are generally high-cost and low-frequency digital traces, which suffer most from the sparsity issue and thus need more supplements from other behavior types. In addition, this result further illustrates the necessity of exploiting multiple behavior types in recommender systems since relying on single strong primary behavior may hard to represent user preferences well. Besides, this pattern is more evident in e-commerce scenarios where the $f_s$ learned is up to 0.91, suggesting that the behavior types of \textit{click, favorite}, and \textit{add-to-cart} play significant roles in the prediction of \textit{purchase} behavior that should not be neglected.

\subsubsection{Embedding Similarities between Behaviors}
\begin{figure}
    \centering
    \includegraphics[width=0.35\textwidth]{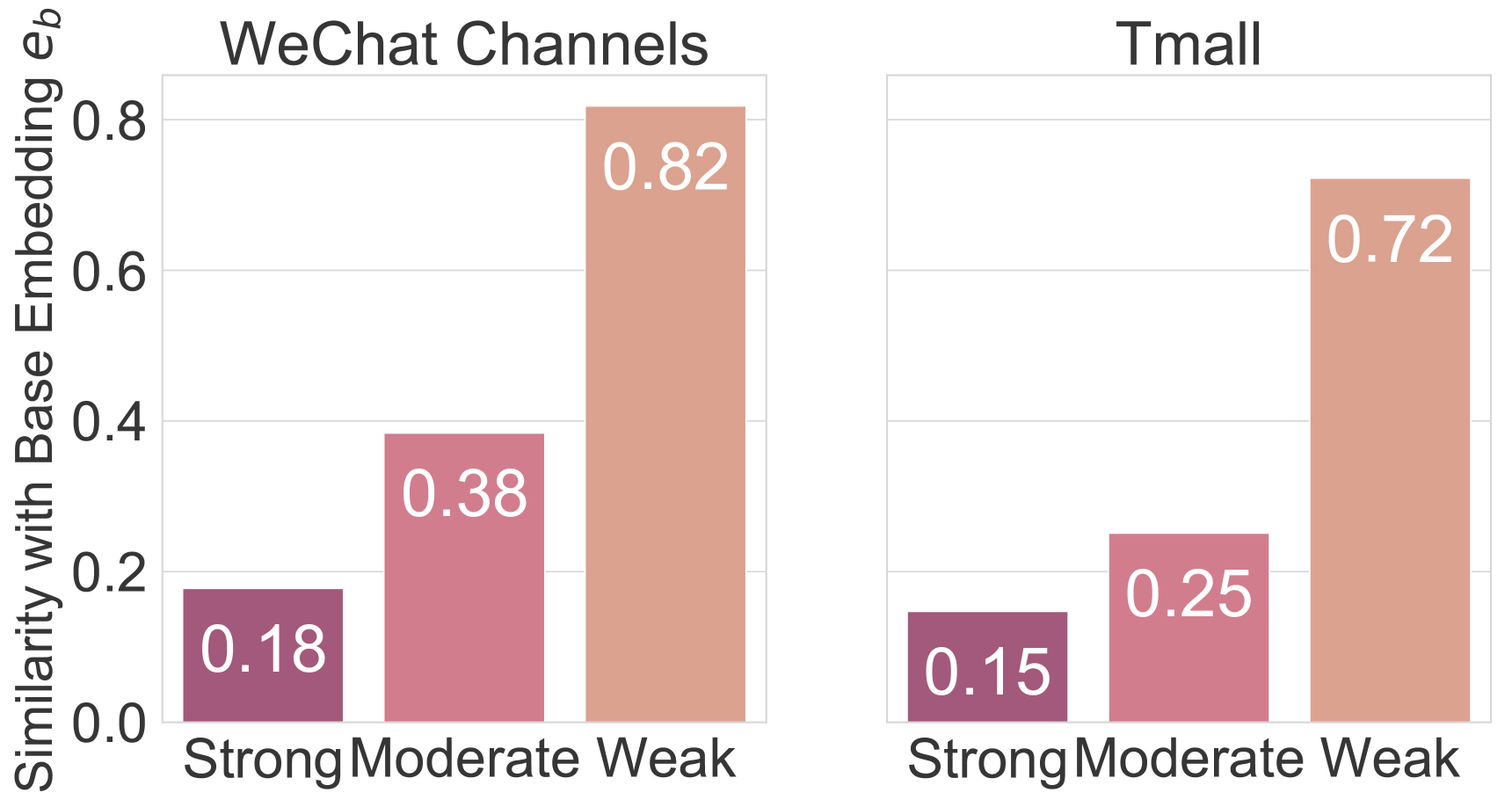}
    \caption{Study of similarities between individual behavior class embeddings and the base embedding $e_b$.}
    \label{fig:Similarity_eb}
\end{figure}

To further investigate the associations among various user feedback types, we also calculate the cosine similarities between the embedding of each behavior category in $\left \{e_s,e_m,e_w \right \}$ and the base embedding $e_b$ on WeChat Channels and Tmall datasets. From the results in Fig. \ref{fig:Similarity_eb}, we can observe that the user representation of \textit{strong} behavior class presents the lowest similarities to the base embedding, followed by the \textit{moderate} and \textit{weak} classes. In addition, the order of behavior classes is consistent between the two datasets. This phenomenon may suggest that the user representations of behavior types with larger amounts of interactions (e.g., \textit{watch} and \textit{click}) tend to be more similar to the base user embeddings regardless of recommendation scenarios. When looking at the specific numeric values, we can find it varies across scenarios that WeChat Channels offers relatively higher values than Tmall. This observation may indicate that user behavior types in short-video applications seem to be more similar to each other, compared to e-commerce platforms.  

\subsection{Optimum Number of Buckets}\label{num_bucket}

To ascertain the optimal number of buckets for behavior classification in \textit{Tricolore}, we conducted comprehensive experiments under various settings and observed that employing three buckets generally yields superior results compared to other configurations. Using WeChat as an illustration, the enhancements achieved with three buckets, as opposed to two, are detailed in Table \ref{tab:bucket_exp}. In the two-bucket classification, \textit{watch} is placed in one bucket, and all other behavior types in the other. Consistently, for the other two datasets employed, the optimal number of buckets is also determined to be three. However, it is crucial to acknowledge that the optimal number of buckets may vary when users apply \textit{Tricolore} to their own data. It is important to clarify that the \textit{Tricolore} framework proposed is a versatile one, and the recommendation of three buckets is not rigidly fixed.

\begin{table}[]
	\caption{Comparison between different number of buckets on WeChat.}
	\label{tab:bucket_exp}
	\centering
	\setlength{\abovecaptionskip}{0.09 cm}
	\setlength{\belowcaptionskip}{0 cm}
	\setlength{\tabcolsep}{2.4 mm}
	\renewcommand\arraystretch{1.3}
	\footnotesize
	\begin{tabular}{c||c|c|c|c}
		\hline 
		&\bf HR@5&\bf HR@10&\bf NDCG@5&\bf NDCG@10\\
		\hline 
        \hline 
		\textbf{Two Buckets} & 0.3109 & 0.4460 & 0.2077 & 0.2508 \\
		\hline 
        \textbf{Three Buckets} & 0.3230 & 0.4737 & 0.2189 & 0.2674 \\
		\hline
        \textbf{Impr.} & +3.89\% & +6.21\% & +5.41\% & +6.62\% \\
		\hline 
	\end{tabular}
	\captionsetup{margin=0cm}
\end{table}

\subsection{Bucket-wise Prediction Task}
Having seen the overall effectiveness of \textit{Tricolore}, we think it is also intriguing to assess its performance in each individual bucket recommendation task. Here we compare the bucket-wise results of \textit{Tricolore} with the best baseline MC-BPR on WeChat data. It can be seen from Table \ref{tab:individual_bucket_exp} that our model offers better results in each individual task. In addition, stronger behaviors present relatively larger improvements. This pattern is also observed on Tmall that the improvements of \textit{Tricolore} in strong behaviors (\textit{purchase}) is the largest, suggesting that the proposed model does not perform well at the cost of performance losses for stronger signals. Instead, for strong behaviors that many conventional single-task recommender systems have concerned most, \textit{Tricolore} performs even better, demonstrating its business values for real-world platforms.

\begin{table}[]
	\caption{Prediction for individual bucket on WeChat data.}
	\label{tab:individual_bucket_exp}
	\centering
	\setlength{\abovecaptionskip}{0 cm}
	\setlength{\belowcaptionskip}{0 cm}
    \setlength{\tabcolsep}{1.5 mm}
	\renewcommand\arraystretch{1.3}
	\footnotesize
	\begin{tabular}{c|c||c|c|c|c}
		\hline 
		{\bf Bucket}& {\bf Methods}&\bf HR@5&\bf HR@10&\bf NDCG@5&\bf NDCG@10 \\
		\hline 
		\hline 
		&\bf MC-BPR & 0.3567 & 0.5158 & 0.2428 & 0.2942 \\
		\cline {2-6} 
        \multirow{-1}*{\cellcolor{Scolor}{\color{white}\bf Strong}}&\bf Tricolore & 0.4261 & 0.5352& 0.2880 & 0.3238 \\
		\cline {2-6}
        &\bf Impr. & +19.40\% & +3.76\% & +18.62\% & +10.07\%\\
		\hline 
		&\bf MC-BPR&  0.3304 & 0.4780 & 0.2234 & 0.2713  \\
		\cline {2-6} 
		\multirow{-1}*{\cellcolor{Mcolor}{\color{white}\bf Moderate}}&\bf Tricolore & 0.336 & 0.4815 & 0.2287 & 0.2755 \\
		\cline {2-6}
        &\bf Impr. & +1.66\% & +0.73\% & +2.4\% & +1.54\%\\
		\hline 
		&\bf MC-BPR &  0.3023 & 0.4523 & 0.2062 & 0.2545 \\
		\cline {2-6} 
		\multirow{-1}*{\cellcolor{Wcolor}{\color{white}\bf Weak}}&\bf Tricolore & 0.3227 & 0.4549 & 0.2206 & 0.2631 \\
  		\cline {2-6}
        &\bf Impr. & +6.73\% & +0.58\% & +6.97\% & +3.37\%\\
		\hline 
	\end{tabular}
	\captionsetup{margin=0cm}
\end{table}

\subsection{Experiments on Cold Start Users}\label{cold}

With the implementation of the shared base embedding strategy in \textit{Tricolore}, designed to assist users with limited interaction history in shaping user representations, we specifically targeted 20\% of users with the fewest interactions from the WeChat dataset for cold-start experiments. The results of \textit{Tricolore} in this cold-start setting are then compared with the outcomes among the MBRS methods, as detailed in Table \ref{tab:coldstart_exp}. Encouragingly, our model demonstrates significant improvements in the cold-start recommendation setting compared to all MBRS baselines across various metrics, highlighting \textit{Tricolore}'s superiority in addressing cold-start problems. In comparison to MMCLR, the improvements across metrics are substantial, exceeding 70\%. Furthermore, concerning the two \textit{HR} metrics, the enhancements of our approach compared to the baselines are more pronounced than the NDCG metrics, indicating the efficacy of our model for candidate generation tasks. These findings underscore that the advantages derived from the design of shared base embedding in MBRS are manifold, as \textit{Tricolore} effectively mitigates both cold-start user issues and sparse behavior type issues simultaneously.

\begin{table}[]
	\caption{Results for cold-start users on WeChat dataset.}
	\label{tab:coldstart_exp}
	\centering
	\setlength{\abovecaptionskip}{0 cm}
	\setlength{\belowcaptionskip}{0 cm}
	\setlength{\tabcolsep}{2.5mm}
	\renewcommand\arraystretch{1.3}
	\footnotesize
	\begin{tabular}{c||c|c|c|c}
		\hline 
		&\bf HR@5&\bf HR@10&\bf NDCG@5&\bf NDCG@10\\
		\hline 
        \hline 
		\textbf{MC-BPR} & 0.2653 & 0.4036 & 0.1764 & 0.2208 \\
		\hline 
  		 \textbf{$\Delta$} &  (+30.91\%) &  (+40.83\%) &  (+13.44\%) &  (+22.51\%) \\
		\hline 
        \textbf{MBGCN} & 0.2125 & 0.3429 & 0.1334 & 0.1754 \\
		\hline
         \textbf{$\Delta$} &  (+63.43\%) &  (+65.76\%) &  (+50.00\%) &  (+54.22\%) \\
		\hline
        \textbf{MMCLR} & 0.1769 & 0.2958 & 0.1177 & 0.1558\\
		\hline 
         \textbf{$\Delta$} &  (+96.33\%) &  (+92.16\%) &  (+70.01\%) &  (+73.62\%)\\
		\hline 
		\textbf{Ours} & 0.3473 & 0.5684 & 0.2001 & 0.2705\\
		\hline 
	\end{tabular}
	\captionsetup{margin=0cm}
\end{table}

\subsection{Trade-off between Popularity and Accuracy}\label{trade-off}

As detailed in Section \ref{popularity_sampling}, the popularity bias can be naturally alleviated through the popularity-balance technique in negative sampling, penalizing the sampling probability assigned to items with higher popularity. To experimentally illustrate its effectiveness, we compare the models' performance in terms of the average popularity of the top-10 item recommendation lists (ARP metric in \cite{abdollahpouri2019managing}). A higher ARP value indicates a more severe popularity bias. The results in Table \ref{tab:ARP_exp} demonstrate that, aided by the popularity-balance technique, \textit{Tricolore} effectively reduces popularity bias. We apply this technique to MC-BPR as well and observe a significant reduction in popularity bias, highlighting the versatility of the strategy in more general MBRS.

\begin{table}[]
	\caption{Popularity bias for smoothness powers on WeChat Channels.}
	\label{tab:ARP_exp}
	\centering
	\setlength{\abovecaptionskip}{0 cm}
	\setlength{\belowcaptionskip}{0 cm}
	\setlength{\tabcolsep}{4.0 mm}
	\renewcommand\arraystretch{1.3}
	\footnotesize
	\begin{tabular}{c||c|c|c|c}
		\hline 
		&\bf{0.0 (w/o)} & \bf 0.5 &\bf 0.75 &\bf 1.0\\
		\hline 
        \hline 
		\textbf{MC-BPR} & 1181.51 & 942.44 & 841.51 & 808.35 \\
		\hline 
        \textbf{Ours} & 1073.29 & 1014.38 & 867.37 & 802.37 \\
		\hline 
	\end{tabular}
	\captionsetup{margin=0cm}
\end{table}

Experiments conducted across various smoothness power settings reveal a discernible trade-off between item popularity and recommendation accuracy, as shown in Table \ref{tab:polularity_exp}. To determine the optimal smoothness power value, we frame the task as a multi-objective optimization problem involving pairs of accuracy metrics (HR@5, HR@10, NDCG@5, or NDCG@10) and a popularity metric (ARP). We first standardize the values of each accuracy metric ($A$) and the popularity metric ($R$) according to:

\begin{equation}
A_{norm}=\frac{A-A_{min}}{A_{max}-A_{min}},
\label{Equ_ACC_Norm}
\end{equation}

\begin{equation}
R_{norm}=\frac{1/R-1/R_{max}}{1/R_{min}-1/R_{max}}.
\label{Equ_ARP_Norm}
\end{equation}

\begin{figure}[t]
    \centering
    \includegraphics[width=0.4\textwidth]{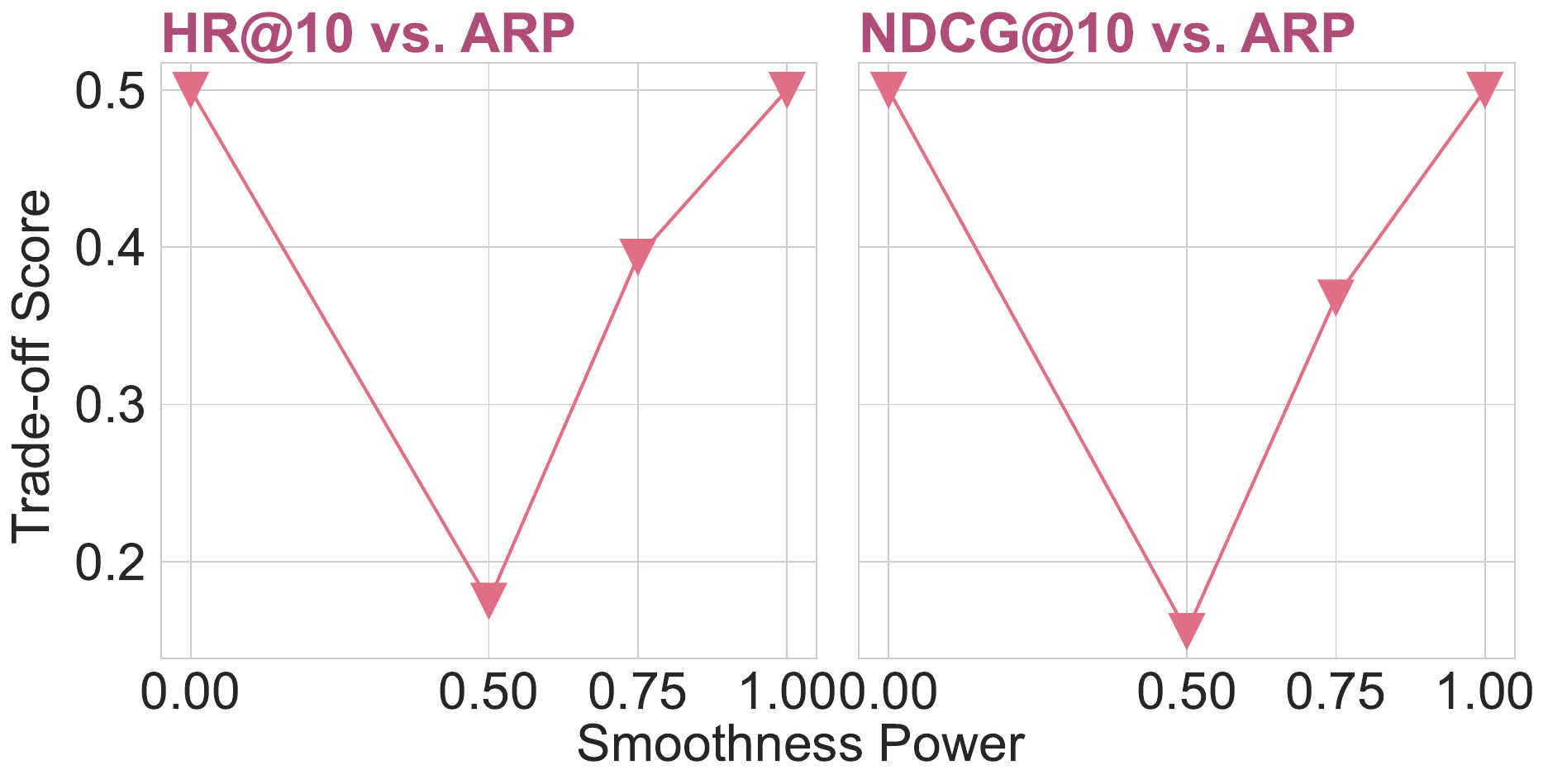}
    \caption{Scoring the Trade-off Between Accuracy and Popularity Metrics.}
    \label{fig:ACC_POP}
\end{figure}

We then compute the trade-off score based on the normalized values of pairs of metrics as follows:

\begin{equation}
t_{s} = \omega_{1} A_{norm}+\omega_{2} R_{norm}, 
\label{Equ_Tradeoff}
\end{equation}
\noindent where $\omega_{1}$ and $\omega_{2}$ denote the weights assigned to the accuracy and popularity aspects, respectively. The choice of specific weight values depends on the platform's prioritization of these two optimization objectives. In our case, we set $\omega_{1}=\omega_{2}=0.5$ and present the corresponding trade-off scores for two accuracy-popularity pairs in the top 10 recommendations in Fig. \ref{fig:ACC_POP}. Notably, a smoothness power of 0.75 consistently yields higher scores, except in the extreme cases of 0 and 1. If smaller powers are employed, the performance of \textit{Tricolore}, as shown in Table \ref{tab:overall_exp}, is expected to improve further.

\begin{table}[]
	\captionsetup{margin=0cm}
	\caption{Results for different smoothness powers on WeChat Channels.}
	\label{tab:polularity_exp}
	\centering
	\setlength{\abovecaptionskip}{0 cm}
	\setlength{\belowcaptionskip}{0 cm}
	\setlength{\tabcolsep}{2.4 mm}
	\renewcommand\arraystretch{1.3}
	\footnotesize
	\begin{tabular}{c||c|c|c|c|c}
		\hline 
		&\bf HR@5&\bf HR@10&\bf NDCG@5&\bf NDCG@10&\bf ARP\\
		\hline 
        \hline 
		\textbf{0.0} & 0.4564 & 0.6134 &  0.3262 & 0.3769 & 1073.29\\
		\hline 
        \textbf{0.5} & 0.3407 & 0.4880 & 0.2318 & 0.2795 & 1014.38 \\
		\hline
        \textbf{0.75} & 0.3230 & 0.4737 & 0.2189 & 0.2674 & 867.37 \\
		\hline 
		\textbf{1.0} & 0.3210 & 0.4608 & 0.2180 & 0.2637 & 802.37 \\
		\hline 
	\end{tabular}
\end{table}

\subsection{Computational Complexity}


Given the importance of computational complexity in candidate generation tasks for MBRS, we compare the time complexity of our model with that of the graph-based MBGCN, the best baseline, MC-BPR, and the classical DSSM candidate generation model.

Let \( N \) denote the number of samples and \( M \) denote the number of layers in the MLP. The computational complexity of \textit{Tricolore} is \( O(M \cdot N \cdot d^2) \). In comparison, the time complexity of MBGCN is \( O(K \cdot E \cdot N \cdot \frac{d}{B}) \), where \( E \) is the number of edges and \( B \) is the batch size. Our model is more computationally efficient because \( \frac{E}{B} \gg d \) and \( E \cdot N \cdot d / B \gg N \cdot d^2 \). The time complexity of MC-BPR is \( O(K \cdot M \cdot N \cdot d^2) \), which is on par with that of \textit{Tricolore}. Compared to DSSM, a popular two-tower model in large-scale recommender systems, \textit{Tricolore} also employs a two-tower structure without cross-interaction training before prediction. The only additional computational cost arises from handling multi-basket behaviors within a multi-task framework. As a result, the computational complexity of \textit{Tricolore} is comparable to that of DSSM.

This analysis demonstrates that while \textit{Tricolore} significantly outperforms existing MBRS models, its computational complexity is equal to or even lower than that of its counterparts.

\section{Discussion}

We introduce a novel approach to mining nuanced preferences from ambiguous feedback. Unlike existing methods that impose rigid temporal or strength constraints on behaviors,
\textit{Tricolore} utilizes a hierarchical representation with base and fine-grained class embeddings. It employs sets of learnable parameters from the initial encoder to the final prediction, facilitating a thorough exploration of behavior associations, thereby making a unique contribution to MBRS research.

However, \textit{Tricolore} has limitations. Currently, it does not address contextual recommendation scenarios like time-aware recommendations, which could enhance accuracy by considering temporal dynamics of user preferences. Moreover, due to the absence of direct negative user feedback in our datasets, modeling negative behavior types has not been emphasized. Yet, we propose refining \textit{Tricolore} by incorporating subtle cues from weak feedback signals, such as short video watching duration or e-commerce click behavior without subsequent actions. Integrating such strategies into the negative sampling module could improve user representation learning. Furthermore, as a future research direction, we aim to integrate social information into the MBRS algorithms, an area largely unexplored by existing models. Many platforms, including the WeChat Channel analyzed in this study, offer social functionalities that could significantly enhance multi-behavior recommendations. For instance, user likes on short videos are visible to their WeChat friends, yet the platform does not share such recommendation reasons for videos watched beyond a certain duration. This design variation reflects different behavior preferences influenced by individual personality or social dynamics, prompting further exploration of multi-task learning frameworks within \textit{Tricolore}.

\section{Conclusion}

We present \textit{Tricolore}, a versatile multi-behavior recommendation framework adept at adaptive learning across diverse behavior types. It hierarchically models comprehensive user preferences during candidate generation, tailored for various recommendation domains. Employing a multi-vector learning approach, it captures distinct behavior characteristics simultaneously. \textit{Tricolore}'s flexible multi-task structure allows customization to specific recommendation needs, augmented by a popularity-balancing technique to mitigate bias. Extensive experiments across public datasets confirm its effectiveness in short videos and e-commerce, particularly excelling in cold-start scenarios. This framework promises to enhance user engagement and recommendation quality significantly.



\bibliographystyle{IEEEtran}
\bibliography{sample-base.bib}




\begin{IEEEbiography}[{\includegraphics[width=1in,height=1.25in,clip,keepaspectratio]{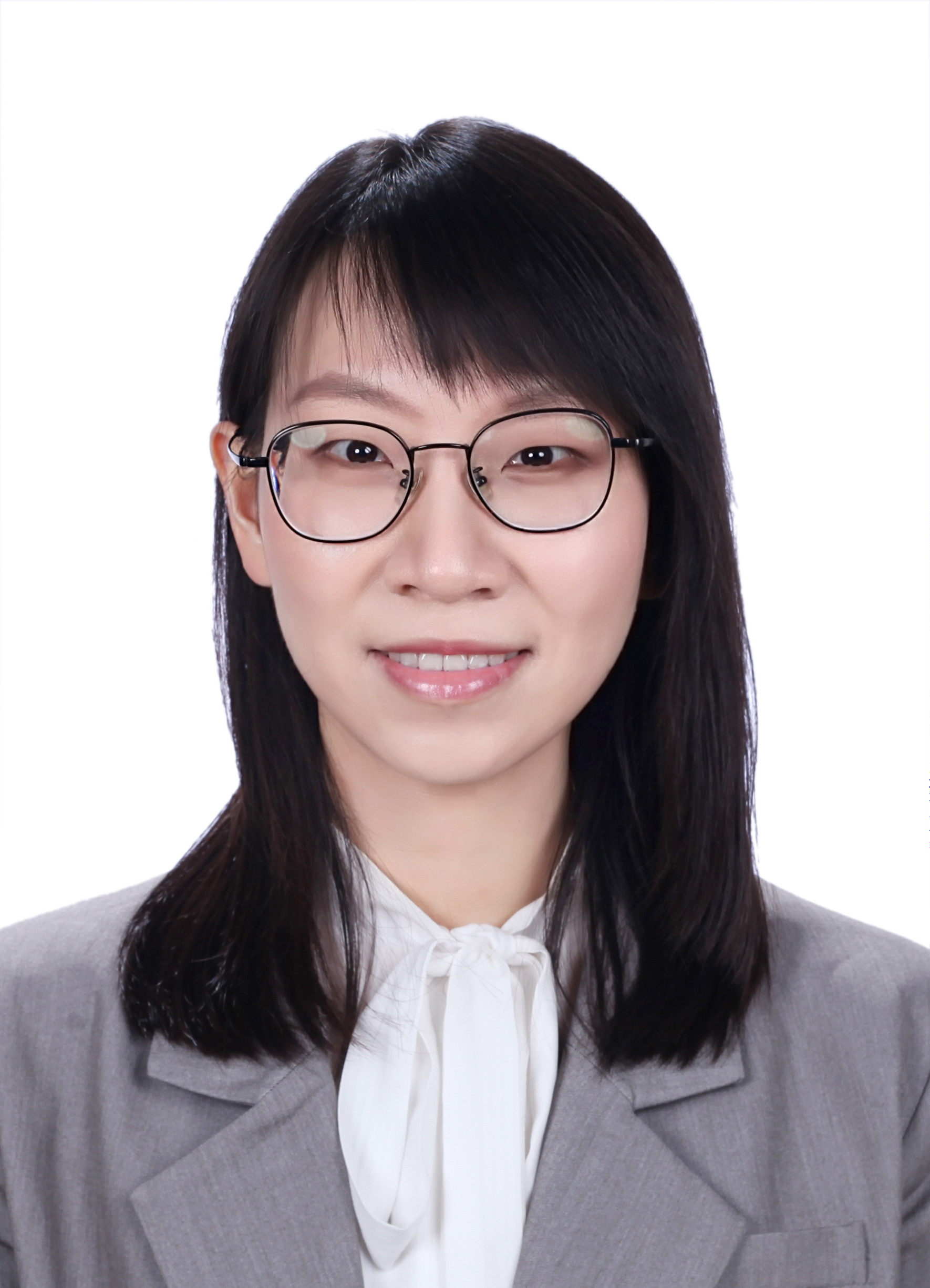}}]{Xiao Zhou}
joined the Gaoling School of Artificial Intelligence at Renmin University of China as a tenure-track Assistant Professor in 2021. Prior to this, she served as a Postdoctoral Fellow at MIT, after completing her Ph.D. in Computer Science at the University of Cambridge, UK. Her research interests include data mining, responsible recommender systems, urban computing, social computing, and large language models. She has expertise in spatio-temporal analysis and is particularly passionate about applying AI to social sciences.

\end{IEEEbiography}

\begin{IEEEbiography}[{\includegraphics[width=1in,height=1.25in,clip,keepaspectratio]{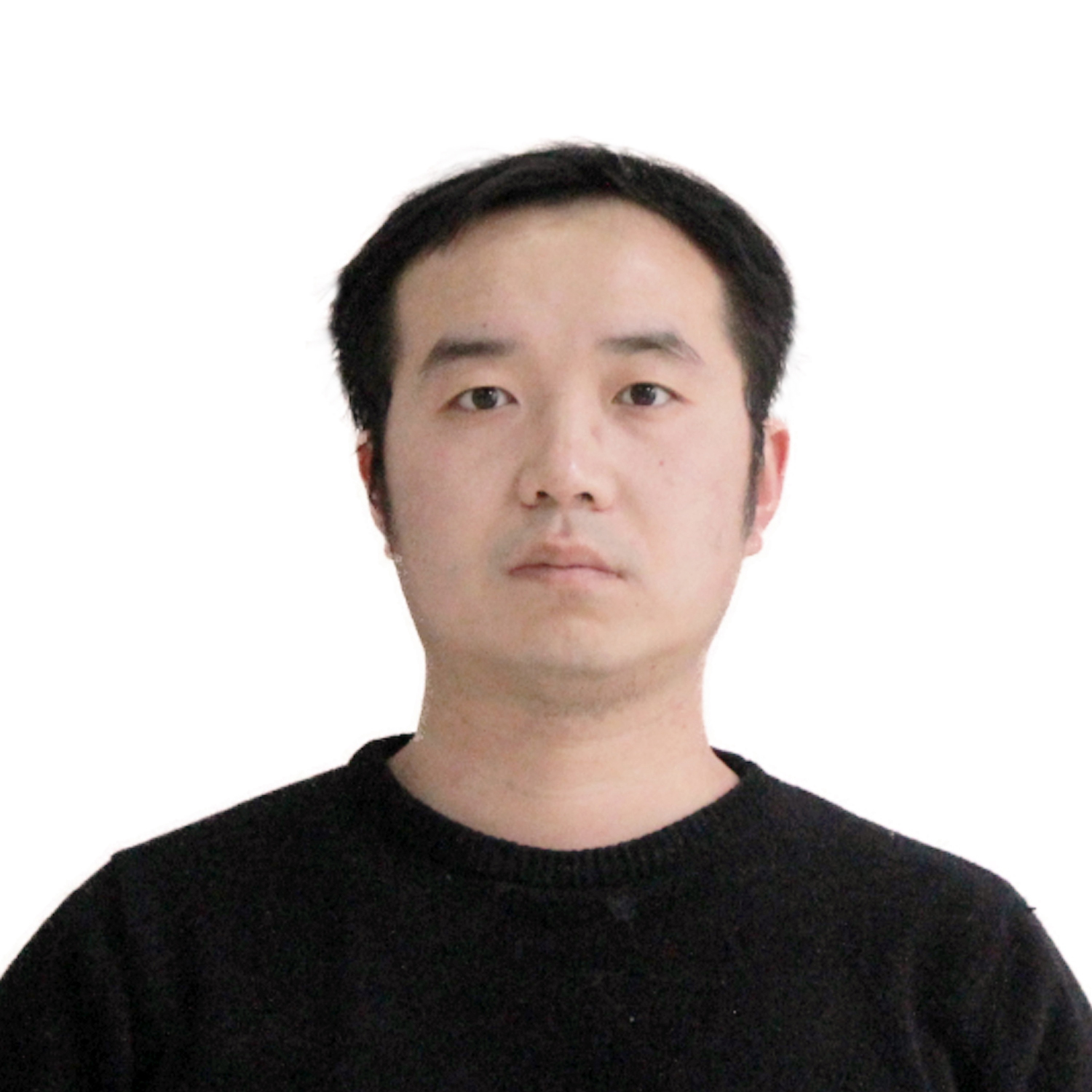}}]{Zhongxiang Zhao} 
is currently a Senior Algorithm Engineer in the WeChat Business Group at Tencent Corporation. He earned his Master's degree in Computer Science from Harbin Institute of Technology in 2016 and has since gained extensive experience at leading technology companies. His expertise focuses on developing recommendation systems for real-world applications, including image-text, short-video platforms, and e-commerce platforms.

\end{IEEEbiography}

\begin{IEEEbiography}[{\includegraphics[width=1in,height=1.25in,clip,keepaspectratio]{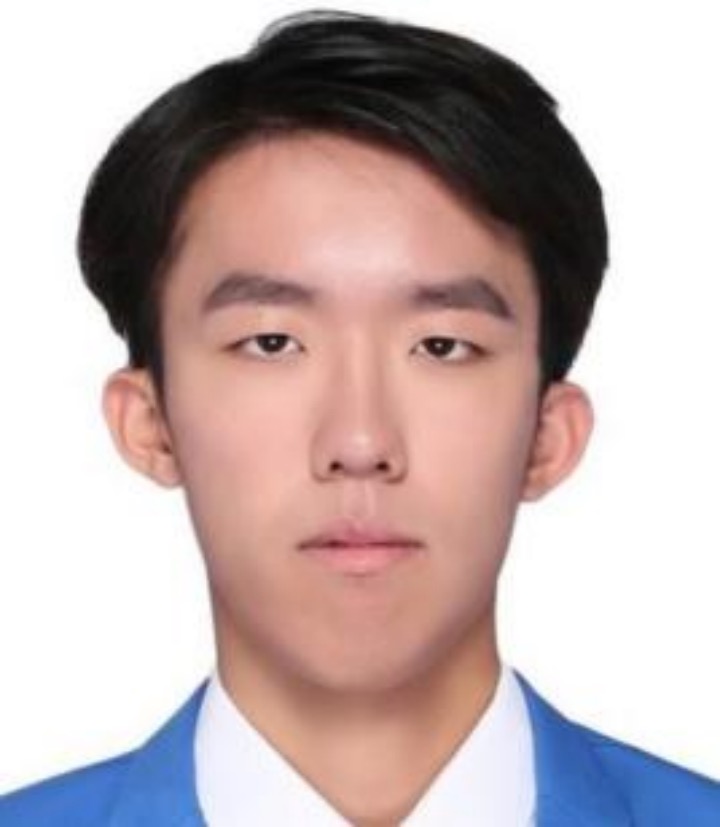}}]{Hanze Guo}
earned his bachelor's degree in Computer Science from the University of Electronic Science and Technology of China in 2022. Currently, he is pursuing his PhD under the supervision of Dr. Xiao Zhou at the Gaoling School of Artificial Intelligence, Renmin University of China. His research interests include deep learning, graph neural networks, and recommendation systems.

\end{IEEEbiography}

\vfill

\end{document}